\documentstyle[twocolumn,aps,psfig]{revtex}
\newcommand{\be}{\begin{equation}}
\newcommand{\ee}{\end{equation}}
\newcommand{\bear}{\begin{eqnarray}}
\newcommand{\eear}{\end{eqnarray}}
 \def\eqbegin {
\begin{eqnarray} } \def\eqend { \end{eqnarray} }
\def\beq{\begin{equation}} \def\eeq{\end{equation}}
 
\def\del{ \partial }

\def\hs_2{\hspace{2mm}}
\def\hs_3{\hspace{3mm}} 

\begin{document}

\twocolumn[\hsize\textwidth\columnwidth\hsize\csname @twocolumnfalse\endcsname
\title{Field Theory of Spin-Singlet Quantum Hall States}

\author{Kazusumi Ino
}                     

\address{Nomura Research Institute,Hongo 2-2-9,Bunkyo-ku,Tokyo,113-0033,Japan}

\maketitle
\begin{abstract}
We formulate a field theory for a class of spin-singlet quantum Hall
states which have been proposed for the quantized Hall plateaus
observed at the second lowest Landau level 
(the Haldane-Rezayi state and its variants).
 A new essential ingredient is  a
class of super Chern-Simons field. We show that the known
properties of the states are consistently described by it. We also give a
$2+1$ dimensional  hierarchical construction. Implications of the proposal 
are discussed and a new physical picture of composite particles
emerges. 
\end{abstract}


]

\narrowtext

{\bf Introduction}
Recently, new phenomena in the high Landau levels of the two-dimensional
electron gas  attract much attention \cite{review}.
Among them is the
quantized Hall plateau found at $\nu=5/2$ , where
$\nu$ is the filling of Landau levels\cite{clark,pan} and  its
denominator  is even.  This is a unique exception to the
odd-denominator rule--the fractional quantized Hall plateaus
are always found around filling fractions with odd-denominators.

Haldane and Rezayi\cite{halrez} proposed an
ansatz wave function for the $\nu=5/2$ state which is
spin-singlet. The tilted field experiments show a collapse of the gap
 in accordance with the proposal of Haldane and Rezayi
\cite{eisen}. Although it has some attractive features as a candidate for
the $\nu=5/2$ plateau, it was argued \cite{MYG} that
the hollow-core model in which the ansatz wave function
 gives the exact ground state is not a
good approximation to the system with Coulomb interaction at the second
lowest Landau levels (N=1 LL)
\cite{LL}.

Numerical studies \cite{gww} of the Coulomb model show
that the ground state at $\nu = 5/2$ may be the spin-polarized
  state proposed by Moore and Read\cite{moore}.
This direction attracts some attention \cite{readgreen,pffield,wen98}.
However, the discussions pose problems in some important respects.

First of all, as this state is spin-polarized,
a new explanation for the results of tilted
field experiments is required.
Until now, some suggestions
such as the effect of thickness of  samples have been  made.
Although the role of spin at N=1 LL is largely unknown,
it has been pointed out that, in specimens to realize
fractional quantum Hall effect (FQHE),
the  effective Zeeman energy can be  so small
 that  the spin degrees of freedom are not frozen out \cite{halp}.
Another  unsatisfactory feature
is in its failure  of  implementing
the idea of hierarchy \cite{hier}.
In view of Jain's successful idea  and
 the prominent role of $\nu=1/2,1/4,\cdots$ composite Fermi liquid
  \cite{hlr} in  FQHE at the lowest Landau level (N=0 LL),
it has been proposed that FQHE at N=1 LL
 may be also dictated by a hierarchy of very different kind \cite{spin}.

The  state also has a  difficulty in  $2+1$ dimensional
field theoretical interpretation.
There have been some proposals of
$2+1$ dimensional field theory all based on
$SU(2)$ Chern-Simons  gauge theory at level 2 \cite{pffield}
(except the one proposed in Ref.\cite{wen98}).
A  problem in these proposals is that they are valid  only at
special unrealistic filling fraction, or
 require an elimination of superfluous degrees of freedom,
 of which a physical meaning is unclear.
This stems from the fact that  the pairing in the  state
is given by two-dimensional  $c=1/2$ Majorana fermion for which
no direct $2+1$ dimensional  interpretation   is known.

For a class of spin-singlet quantum Hall states,
 namely permanent quantum Hall states (variants of 
the Haldane-Rezayi state\cite{YGM}) are identified
as hierarchical states formed on the Haldane-Rezayi state\cite{spin}.
These states inherit most  features of the HR state.
Indeed, the recent experiments at the plateaus around $\nu=5/2$
($\nu=7/3,8/3 ...)$   suggest
a close relation between these plateaus\cite{pan}.

In this paper, we  propose a 2+1 dimensional field theory of
the Haldane-Rezayi state and its variants (permanent quantum Hall states).
 A new essential ingredient is  a
class of super Chern-Simons field. We show that the known
properties of the states are consistently described by it. We also give a
2+1 dimensional  hierarchical construction. Implications  of the proposal 
are discussed 
and new physical picture of composite particle at N=1 LL emerges.

{\bf  Permanent quantum Hall state}
For  the FQHE at N=0 LL, wave functions were given by  conformal blocks
of 2 dimensional conformal field theory in the bulk (${\rm CFT}_2$) while
edge excitations were described by 1+1 dimensional theory on the edge
(${\rm CFT}_{1+1}$).  There ${\rm CFT}_2$ and ${\rm CFT}_{1+1}$
were equivalent and the global symmetry of CFT  was
generated by the gauge symmetry of 2+1 dimensional CS theory.
These are all  based on the relation between the CS theory and
 CFT \cite{witten}.
We expect that this scheme is at work also for N=1 LL, but with
one extension. In the HR or  permanent QH states,
${\rm CFT}_2$ and ${\rm CFT}_{1+1}$ are slightly different although
their  space of states coincides and  the partition functions are
equal with
an inclusion of non-trivial  flux \cite{milo,gfn,iHR}.
This flux  is  only accounted for
 by the twist of the boundary condition in ${\rm CFT}_{1+1}$.
Thus the field content of the 2+1 dimensional theory
will respect the symmetry of ${\rm CFT}_2$.
We shall use this as a guiding principle for constructing
a proper 2+1 dimensional field theory.
Especially  the boundary action deduced from the 2+1
dimensional theory  should be the one for ${\rm CFT}_2$.

Let us recall ${\rm CFT}_2$  of
the $\nu =1/q \hspace{2mm}(q =p+1,\hspace{2mm} p $:even integer)
  permanent QH state.
As  observed in \cite{moore}, the pairing part of the permanent state can be
written as a conformal block of $c=-1$
bosonic ghosts $\beta$-$\gamma$ \cite{FMS}.
In terms of these fields and a chiral boson $\varphi$,
the electrons in the permanent state are represented by
$\beta e^{i\sqrt{q}\varphi},\hspace{3mm}\gamma e^{i\sqrt{q}\varphi}$.
There is a  bosonization  scheme
for the $\beta$-$\gamma$ system \cite{FMS,spin}
in which $\beta$ and $\gamma$ are bosonized into
$c =-2$ symplectic fermion system and a chiral boson $\phi$
with negative signature as follows:
\eqbegin
\beta=\del\theta^{\uparrow}e^{ \phi}, \hspace{5mm}
\gamma=\del\theta^{\downarrow}e^{-\phi}.
\label{bonzo}
\eqend
$\phi$ and $\varphi$   may be generated by U(1) CS fields
in 2+1 dimensional theory.
For the  $c=-2$ part,
there is  a Parisi-Soulas supersymmetry,
which indicates an emergence of  a fermionic gauge symmetry
in 2+1 dimensions.

These considerations lead us to consider  the generators
of gauge symmetry $P^{\mu}$,$Q,\widetilde{Q} \hspace{2mm} (\mu=0,1)$ where
$P^{\mu}$ is for U(1) symmetry, $Q$ and $\widetilde{Q}$
are for possible supersymmetry.
We assume that the U(1) gauge symmetries commute
$[P^{\mu},P^{\nu}] =0$.
To construct a proper commutation relations involving
$Q$ and $\widetilde{Q}$ one must  take into account
an important observation in  Ref.\cite{spin}
that the  formation of filled LL for composite fermions  should
affect the pairing.
A simple way to implement it is to consider
$Q$ and $\widetilde{Q}$ forming a representation of $P^{\mu}$. However,
the effect of filled LL on pairing is noninvertible in the sense that
the converse effect that
filled LL affects unpaired composite fermions to form a pair do not occur.
But the converse effect is
inevitable when $Q$ and $\widetilde{Q}$ form a representation.
To avoid it, we may take $\widetilde{Q}$ as a fermionic central
extension of an ordinary supersymmetry algebra as follows:
\eqbegin
[P^{\mu},P^{\nu}] =0, \hspace{2mm}
\{Q,Q \} = 2\gamma_{\mu}P^{\mu},\hspace{2mm}
[P^{\mu},Q] =2i\gamma^{\mu }\widetilde{Q},
\label{algebra}
\eqend
where we set
$\gamma^0 =\gamma^1 =-\gamma_0=\gamma_1= 1$. Other commutatators
are all set to  zero.
These commutation relations satisfy the Jacobi identity and
form a graded Lie algebra.
We may define an invariant metric by the trace
\eqbegin
{\rm Tr}(P^{\mu}P^{\nu}) =-\frac{1}{2}\eta^{\mu\nu}, \hspace{2mm}
{\rm Tr}(\widetilde{Q}Q) = \frac{i}{2}.
\label{trace}
\eqend
In (\ref{algebra}), supercharges appear unsymmetrically.
To treat them in a symmetric way, we may
take a basis in spin by
$Q^{\uparrow}=\frac{1}{\sqrt{2}}(Q+\widetilde{Q}),
\hspace{2mm} Q^{\downarrow}=\frac{1}{\sqrt{2}}(Q-\widetilde{Q})$.
Similar enlarged gauge symmetry has been  known in
 supergravity \cite{dauria-fre} and also studied in
 the relation between CS theory and superstring theories
 \cite{green}.

The one-form gauge field has the  expansion
\eqbegin
{\cal A}  = iA^{\mu}P_{\mu} + \psi Q
-\widetilde{\psi}\widetilde{Q},
\eqend
where $A^{\mu}$ is  U(1) gauge field,
$\psi$ and $\widetilde{\psi}$ are real fermionic gauge fields.
We  consider the CS action for ${\cal A}$ ( with $k=1$ ),
\eqbegin
S &=& \frac{k}{4\pi}\int {\rm Tr}(  {\cal A}d{\cal A} + \frac{2}{3} {\cal
A}^3) \nonumber \\
 &=& \frac{k}{4\pi}\int  \frac{1}{2}A^{\mu} \wedge d A_{\mu}
+ i \widetilde{\psi}\wedge d \psi
+ i\psi\wedge\gamma^{\mu}\psi\wedge A_{\mu}.
\label{action}
\eqend
The coupling to a source is
achieved by the addition of a term ${\cal J}\wedge {\cal A}$.
The field equation is $d{\cal A} + [{\cal A},{\cal A}] ={\cal J}$,
which is in components
\eqbegin
dA^{\mu} &+&i\psi\wedge \gamma^{\mu}\psi = J^{\mu},
\label{fieldeq1}  \hspace{4mm}
d(\psi_{\uparrow} + \psi_{\downarrow}) = J_{Q},
\label{fieldeq2}\\
d(\psi_{\uparrow} &-& \psi_{\downarrow})
-2(\psi_{\uparrow} + \psi_{\downarrow})\gamma^{\mu} \wedge A_{\mu} =
J_{\widetilde{Q}}.
\label{fieldeq3}
\eqend
The third equation tells that when two sources with fermions
with opposite spins get closer,
they will be deflected by the bosonic gauge fields.
This is a kind of dynamics supposed to happen in the  permanent states.

The action has the gauge symmetry
$\delta{\cal A}=d\Lambda +[{\cal A}, \Lambda] $ on a closed space.
However on a space with boundary such as cylinder, it is necessary
to add a boundary action.
We will now show that it actually coincides with
the action of ${\rm CFT}_2$ of the permanent state.
To derive it, we note that
the field equation with no source is
solved by   ${\cal A}_{\alpha} ={\cal U}^{-1}\del_{\alpha} {\cal U}
,\hspace{2mm} {\cal U} = {\rm exp}(X),\hspace{2mm}
X =i\varphi^{\mu}P_{\mu}+i{\theta}Q-i\widetilde{\theta}\widetilde{Q}$.
In components,  $A^{\mu}_{\alpha}=\del_{\alpha} \varphi^{\mu}
-i\theta\gamma^{\mu}\del_{\alpha}\theta,
\psi_{\alpha}=\del_{\alpha}\theta,\hspace{1mm}
\widetilde{\psi}_{\alpha}
=\del_{\alpha}\widetilde{\theta}+\gamma^{\mu}\theta\del_{\alpha}\varphi_{\mu}-
\gamma^{\mu}\del_{\alpha}\theta\varphi_{\mu}$.
By using the solution, we get the boundary action for the cylinder
($g_{+-}=g_{-+}=-1$)
\eqbegin
S_{\rm B} &=& -\int \large\{A^{\mu}_{+}A_{-\mu}
+i\psi^{\alpha}\widetilde{\psi}_{\alpha}\large\}d^2\sigma
+ i \int \epsilon^{\alpha\beta\gamma}\psi_{\alpha}
\gamma_{\mu}\psi_{\beta} A^{\mu}_{\gamma} d^3\sigma \nonumber \\
&=&-\int \large\{ (\del_{+}\varphi^{\mu}-i\theta\gamma^{\mu}\del_{+}\theta)
(\del_{-}\varphi_{\mu}-i\theta\gamma_{\mu}\del_{-}\theta)  \nonumber \\
&+&
i\del^{\alpha}\theta
[\del_{\alpha}\widetilde{\theta}+\gamma^{\mu}\theta\del_{\alpha}\varphi_{\mu}
-\gamma^{\mu}\del_{\alpha}\theta \varphi_{\mu}
] \nonumber
-i \epsilon^{\alpha\beta}\del_{\alpha}\varphi^{\mu}
\theta \gamma_{\mu}\del_{\beta}\theta
\large\}d^2\sigma.
\label{Baction1}
\eqend
Calculation shows that some terms including four fermi terms cancel.
Then we have
\eqbegin
S_{\rm B} =-\int \large\{ \del_{+}\varphi^{\mu}
\del_{-}\varphi_{\mu} +
i\del^{\alpha}\theta\del_{\alpha}\widetilde{\theta}
-i \epsilon^{\alpha\beta}\del_{\alpha}\varphi^{\mu}
\theta \gamma_{\mu}\del_{\beta}\theta
\large\}d^2\sigma.
\label{Baction2}
\eqend
In (\ref{Baction2}), we notice that a field redefinition of
$\widetilde{\theta}$ by $\widetilde{\theta}+\theta F$ generates
a term similar to the last one. By taking $F$ satisfying
$\del_{\alpha}F=\gamma^{\mu}\epsilon_{\alpha\beta}\del^\beta \phi_{\mu}$,
we end up with
\eqbegin
S_{\rm B} = -\int \{\del_{+}\varphi^{\mu}
\del_{-}\varphi_{\mu} + i\del^{\alpha}\theta
\del_{\alpha}\widetilde{\theta} \} d^2\sigma.
\label{Baction3}
\eqend
This is   the action for ${\rm CFT}_2$ of the $\nu=1$ permanent state.
For the $\nu=1/q$ state, we  add additional CS term
$\frac{p}{2}A^{1} \wedge d A^{1}$ to the Lagrangian.
This is because $A^1$ gives rise to the Laughlin-Jastrow factor for
filled LL
in the ground state wave function which, for the Laughlin state,
 comes from the CS  field for flux attachment.

The field redefinition to get the action (\ref{Baction3})
spoils the explicit relation between the local supersymmetry
and the global supersymmetry
$Q^{a}_{\rm global}, P^{0}_{\rm global},P^{1}_{\rm global}$.
But the correspondence
 can be given   by characterizing representations by
the charges under two U(1) symmetries and the fermions.

Thus we have shown that the CS theory based on the algebra (\ref{algebra})
leads to the correct ${\rm CFT}_2$ of the permanent QH state.
This implies that it is a proper field theory for the state.
 For ${\rm CFT}_{1+1}$, we should take the
 corresponding unitary theory which has the same  partition function as
the ${\rm CFT}_2$ \cite{spin}.

{\bf (Quasi-) Particles from the Wilson lines }
In Ref.\cite{read95}, it was realized that
 earlier concept of flux attachment \cite{zhang}
 is promoted to vortex attachment.
This is natural from the CS field theoretic
point of view since  composite (quasi-)particles can now be
expressed by the Wilson lines.
This is readily extended to our case.

The Wilson line in our system is
\eqbegin
W = {\rm Tr}_{\cal R} P{\rm exp}(\int_{C} \cal A)
\eqend
where $P$ indicates the path-ordered product, $C$ is a contour
and ${\rm Tr}_{\cal R}$ is the trace for a representation of
${\cal R}$ of the source.
 Wave functions are obtained as an  expectation value in the
presence of  $W$ with the position braket at a time slice.
For  $N$ Wilson lines, their current is
a sum of delta functions at their contours
${\cal J} = \sum_{i=1}^{N} j^s T_s \delta(C_i) $
where $T_s$ represents the generators of the gauge group.
We identify the species of Wilson lines by
their representation under $Q^{a}_{\rm global},
P^{0}_{\rm global}$ and $P^{1}_{\rm global}$,
denoting them by the fermions and the charges as
$(\psi^{a}_{s},q^0,q^1)$.
Here $s$ represents the holonomy for the fermion.

First of all, we consider
$(\psi^{\uparrow}_{\rm R},1,q),(\psi^{\downarrow}_{\rm R},-1,q)$
where $R$ represents the trivial holonomy.
From the relation between CS theory and the ${\rm CFT}_2$,
these Wilson lines correspond to
$\del \theta^{\uparrow}e^{\phi}e^{i\sqrt{q}\varphi},
\del \theta^{\downarrow}e^{-\phi}e^{i\sqrt{q}\varphi}$
i.e.   composite fermions in the system.
The ground state of the permanent QH state for
$2N$ electrons is obtained as
 a collection of $N$ $(\psi^{\uparrow}_{\rm R},1,q)$ and
$N$ $(\psi^{\downarrow}_{\rm R},-1,q)$
(with uniform background field for $A^{1}$).

Let us next consider quasiparticles.
For the paired state, the elementary
quasiparticle will appear with flux halved.
It is a configuration of $A^1$ with a holonomy
$e^{i\pi}$.
Also such quasiparticle generates a non-trivial
phase shift for $A^{0}$ or $\psi^{a}$. As shown in
Ref.\cite{spin}, the consistency of the edge theory
admits the existence of such a configuration
 only for $\psi^{a}$.
For   $\psi^{a}$, the holonomy in the configuration is
$e^{i\pi}$.
We denote this species as $\psi_{\rm NS}$.
Thus the elementary quasiparticle and  quasihole are
 characterized as the species $(\psi^{a}_{NS},0,-1)$,
$(\psi^{a}_{NS},0,1)$.

It is possible to characterize the paired condensate
by gauge symmetry breaking.
 For FQHE at  N=0 LL,  U(1) gauge symmetry is
spontaneously broken by the condensation of composite boson \cite{zhang}.
In our case, two species of composite fermions condensate.
The  super gauge symmetry is divided into two parts one of which
vanishes in the configuration of one species of composite fermion.
A collection of one species   breaks half
of the super gauge symmetry.
When both species are  present, both parts of
the super gauge symmetry are broken.

We also note that half of the supersymmetry which
doesn't vanish in the configuration of one species of composite fermion
produce a fermionic zero mode in the configuration.
The configuration with  a fermionic zero mode gives the remaining
species of quasiparticle in the system.
It  corresponds to the logarithmic field in the ${\rm CFT}_2$.

{\bf Hierarchy}
The construction we have considered can be extended to
hierarchical scheme. As in FQHE at N=0 LL, we introduce  $m$ kinds of flux
described by $m$  U(1) CS gauge field
$A^l  (l=1,\cdots,m).$
The couplings between $\varphi_i$ are specified by
an integer matrix $K$ \cite{read90,frohlich}.
The Jain's hierarchy is
obtained by taking  $K$ as $
K_{ln} = \pm \delta_{ln} +  pC_{ln} $
where $p$ is an  even integer and $C_{ln} =1 $ for
$l,n =1, \cdots,m$.
The filling fraction is given by
$\nu= m/(mp \pm 1) $.
As shown in \cite{frohlich}, this system of U(1)
CS fields is equivalent to the system of SU(m) CS
 field at level 1 and U(1) CS field at level $1/{\nu}$.
As above, we extend this  U(1) symmetry to (\ref{algebra})
by introducing  fermionic CS fields $\psi$, $\widetilde{\psi}$
and bosonic field $A^{0}$.
The composite fermions are expressed by the Wilson lines
of $(\psi^{\uparrow},{\bf 1},1,1/{\nu}),
(\psi^{\downarrow},{\bf 1},-1,1/{\nu})$
 where ${\bf 1}$ stands for the fundamental representation of SU(m).
The quasiparticles under the presence of the collection
of composite fermions reproduce the ones given in \cite{spin}.

{\bf The Haldane-Rezayi state}
Let us turn to the $\nu=1/p$ HR state.  The permanent state is formed by
integer QHE (IQHE) on the HR state.  Then, for the HR state, one expect that
$Q$ and $P$ may be decoupled since
no LL for composite fermion is formed there.
This may be simulated by scaling $P$ to zero.
However it also requires the vanishment of $\widetilde{Q}$ through
the algebra (\ref{algebra}), thus
does not lead to a sensible construction.

Actually alternative viewpoint is possible: the HR state is
a permanent QH state with IQHE turned off.
This results in an algebraic relation of wave functions.
It is known that the determinant of the HR wave function and
the permanent factor of the permanent state wave function
have a simple relation \cite{halrez}
\eqbegin
{\rm det}\left( \frac{1}{(z^{\uparrow}_i-z^{\downarrow}_j)^2}\right) =
{\rm per}\left( \frac{1}{z^{\uparrow}_i-z^{\downarrow}_j}\right)
{\rm det}\left( \frac{1}{z^{\uparrow}_i-z^{\downarrow}_j}\right).
\eqend
Here ${\rm det}\left( \frac{1}{z^{\uparrow}_i-z^{\downarrow}_j}\right)$
is interpreted as the factor to `turn off` IQHE.
It is the same as   the pairing part of the 331 state \cite{halp}, and thus
arises from the U(1) CS theory at level 1. Its generator actually
coincides with $P^1$ in the algebra (\ref{algebra}).
For the permanent QH state, we also took $P^1$ to implement
the even number of flux for composite fermion. For the HR state,
we do not do so not to change the level of the U(1) CS gauge field.
Thus we take the algebra (\ref{algebra}) for the gauge symmetry of
the pairing part only, and take another U(1) CS field $A^2$
at level $p$ for the flux of composite fermion.
Composite fermions are given by the Wilson lines of
$(\psi^{\uparrow}_{R},1,1,p)$ and $(\psi^{\downarrow}_{R},-1,-1,p)$

At first sight, it does not correspond to
the known $c=-2+1$ ${\rm CFT}_2$ of the HR state
since it contains additional two bosons for pairing part
and gives $c=0+1$.
Evidences that it actually does describe the HR state are in order.

First the chiral operator product algebra (OPA) generated by
the CS theory  coincides with the known OPA of
the HR state. Since the OPA determines the correlation between
quasiparticles, their statistics and  topological
order, it implies that the proposed theory is describing the HR state.
Also the cylinder partition function generated by the OPA using
the fermionic character formula agrees with the one for ${\rm CFT}_2$
except for the contribution from the central charge.

Second evidence is concerned with the central charge.
In Ref.\cite{witten},
it was shown that the so-called `flaming` dependence of
the phase of the partition function of a CS  theory
determines the central charge. For convenience, we take
the large $k$ limit and consider one-loop calculation.
It involves the determinant of the fermions. The proposed model
has a nice feature that
the fermions couples to  the bosonic CS fields off-diagonally,
thus the absolute value  of the fermion determinant and
the determinant from boson cancel in the pairing part.
Thus the flaming dependence of the pairing part comes only from the phase
of the fermion determinant
and  gives  $c=-2$.
This value of the central charge is also necessary
to achieve the modular invariance of the cylinder partition
function by the twisting mechanism of Ref.\cite{iHR},
since, otherwise, each part of the partition function
wouldn't transform linearly under the modular transformations.
This indicates the necessity of a correction to the naive counting of
the central charge in  the extended version of the CS-CFT
correspondence.
For the permanent state, half of the phase dependence
is shown to be  canceled by the boson determinant, and thus
the flaming dependence  tells that it has $c=-1+1$.
These calculations are expected to be one-loop exact by virtue of
supersymmetry.
The flaming dependences also indicate the consistency of our proposal.

In this theory, spin degrees of freedom interact with
the U(1) CS  field.
It is expected as in Ref.\cite{spin}
for the permanent QH states, but is
unexpected for the HR state. The algebra (\ref{algebra}) may be
intrinsic to spin-singlet pairing.

\begin{figure}
\vspace{5.0cm}
\centerline{\psfig{figure=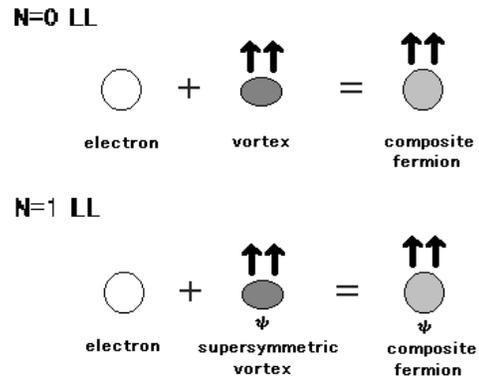,width=7.5cm,angle=0}}
\vspace{0.2cm}
\caption{
At N=0 LL, composite fermion is formed by electron and vortex.
At N=1 LL, composite fermion may be formed by electron
and supersymmetric vortex.
}
\label{CF}
\vspace{0.2cm}
\end{figure}

We remark that the characterization of the paired condensate
via spontaneous breaking of the super gauge symmetry  is valid also
for the HR state.  Accordingly the concept of composite particle may be
modified.
At $\nu=1/2$, composite fermion is formed by electron and vortex
while, at $\nu=5/2$, composite fermion may be
formed by electron and supersymmetric vortex (Fig.{\ref{CF}}).

{\bf Discussions}
We add a matter to our theory by the Wilson lines,
which may be suitable for an application to knot theory.
It is also possible to couple a second-quantized matter by
 the supermultiplet containing complex boson while
 respecting the local symmetry.
We plan to address a precise formulation in a subsequent paper.
Such a formulation enables a quantitative
study of the dynamical mechanism of pair formation while
qualitative  behavior can be deduced within our approach as
we now describe.

Perturbative treatment of quasiparticle at the $\nu=1/2$ state \cite{hlr}
 is  justified  by the renormalization group
flow of a scaling onto the Fermi surface \cite{nw}
 with the proviso of renormalization of the effective mass.
Based on this, in Ref.\cite{bonesteel}, it was shown
that the U(1) CS  field  is highly pair breaking.
For the Coulomb interaction, one must resort to a softening
of the interaction by the thickness of real specimen
for pairing to occur.


Now let us apply the same method  to our case.
In our case, strong fluctuation of the U(1)  CS gauge field
for a spin-singlet pair  is  suppressed
by the contribution from the fermionic gauge field for spin.
Thus the pair breaking effect is largely
tamed for the spin-singlet pair.
This reduction  of pair breaking effect of composite fermions
 suggests a similar mechanism at the microscopic level.
Actually an example of such mechanism is already known :
the HC model \cite{halrez}.
There the pairing is induced by the reduction of
zero-th order pseudopotential between electrons.
Our observation  implies that,
if a HC-like model is really valid  at N=1 LL,
it should involve spin as its essential cause.
Indeed, tilted field experiments have been suggesting that
 the spin degrees of freedom may be important at  N=1 LL.

Finally it may be interesting to apply our construction to
similar planer systems such as given in Ref.\cite{bfn,sf}.
We plan to address this issue \cite{subseq}.

{\it Acknowledgement} The author is  grateful to
Mahito Kohmoto   for  useful discussions 
and to Takashi Nakano at RCNP of Osaka University 
for correspondence.

\vskip 0.2in
\noindent

\def\NP{{Nucl. Phys.\ }}
\def\PRL{{Phys. Rev. Lett.\ }}
\def\PL{{Phys. Lett.\ }}
\def\PR{{Phys. Rev.\ }}
\def\IJMP{{Int. J. Mod. Phys.\ }}

\begin{thebibliography}{99}
\bibitem{LL} The Landau levels are classified by N. The correspponding
noninteracting electronic energy levels are $(N+1/2)\hbar \omega_c$ where
$\omega_c$ is the cyclotron frequency $eB/m_e c$. $B$ is the magnetic field and
$m_e$ is the effective mass of electrons. The lowest Landau level is
$N=0$ LL and the second lowest Landau level is $N=1$ LL. Since electrons
have spin
degree of freedom, each Landau level has $\nu=2$ if it is totally filled by
electrons with up and down spins. (The Zeeman splitting is neglected here.)
\end{thebibliography}
\end{document}